\documentclass{PoS}

\usepackage{axodraw}
\usepackage{acronym}
\usepackage{multirow}

\acrodef{SM}{Standard Model}
\acrodef{EW}{electroweak}
\acrodef{LO}{leading order}
\acrodef{NLO}{next-to-leading order}
\acrodef{NNLO}{next-to-next-to-leading order}

\def\nl{\nonumber\\}

\def\beq{\begin{equation}}
\def\eeq{\end{equation}}
\def\beqar{\begin{eqnarray}}
\def\eeqar{\end{eqnarray}}
\def\barr#1{\begin{array}{#1}}
\def\earr{\end{array}}
\def\bfi{\begin{figure}}
\def\efi{\end{figure}}
\def\btab{\begin{table}}
\def\etab{\end{table}}
\def\bce{\begin{center}}
\def\ece{\end{center}}

\def\text{\textstyle}

\def\al{\alpha}

\def\ga{\gamma}
\def\de{\delta}

\def\De{\Delta}

\def\refeq#1{\mbox{(\ref{#1})}}
\def\reffi#1{\mbox{Fig.~\ref{#1}}}

\def\refta#1{\mbox{Table~\ref{#1}}}

\def\citere#1{\mbox{Ref.~\cite{#1}}}
\def\citeres#1{\mbox{Refs.~\cite{#1}}}

\newcommand{\GeV}{\unskip\,\mathrm{GeV}}

\newcommand{\TeV}{\unskip\,\mathrm{TeV}}
\newcommand{\fba}{\unskip\,\mathrm{fb}}

\def\mathswitch#1{\relax\ifmmode#1\else$#1$\fi}
\def\mathswitchr#1{\relax\ifmmode{\mathrm{#1}}\else$\mathrm{#1}$\fi}
\def\mathswitchit#1{\relax\ifmmode{#1}\else$#1$\fi}

\newcommand{\PW}{\mathswitchr W}
\newcommand{\PZ}{\mathswitchr Z}
\newcommand{\Pg}{\mathswitchr g}
\newcommand{\PH}{\mathswitchr H}
\newcommand{\Pd}{\mathswitchr d}
\newcommand{\Pj}{\mathswitchr j}
\newcommand{\Pl}{\mathswitch l}
\newcommand{\Pu}{\mathswitchr u}
\newcommand{\Ps}{\mathswitchr s}
\newcommand{\Pb}{\mathswitchr b}
\newcommand{\Pc}{\mathswitchr c}

\newcommand{\Pp}{\mathswitchr p}
\newcommand{\Pq}{\mathswitchit q}

\newcommand{\Plp}{{\Pl^+}}
\newcommand{\Plm}{{\Pl^-}}
\newcommand{\MW}{\mathswitch {M_{\PW}}}

\newcommand{\MZ}{\mathswitch {M_{\PZ}}}

\newcommand{\GW}{\mathswitch {\Gamma_{\PW}}}
\newcommand{\GZ}{\mathswitch {\Gamma_{\PZ}}}

\newcommand{\GF}{\mathswitch {G_\mu}}

\newcommand{\alphas}{\alpha_{\mathrm{s}}}

\newcommand{\pT}{p_{\rm T}}

\def\ie{i.e.\ }

\def\ea{et al.}

\newcommand{\ord}{{\cal O}}
\renewcommand{\O}{{\cal O}}

\newcommand{\ri}{{\mathrm{i}}}

\newcommand{\rT}{{\mathrm{T}}}

\newcommand{\rd}{{\mathrm{d}}}

\newcommand{\M}{{\cal {M}}}

\newcommand{\EW}{{\mathrm{EW}}}

\newcommand{\LO}{{\mathrm{LO}}}
\newcommand{\NLO}{{\mathrm{NLO}}}

\newcommand{\recola}{{\sc Recola}}
\newcommand{\collier}{{\sc Collier}}

\title{Electroweak corrections to Z + 2 jets production at the LHC }

\ShortTitle{Electroweak corrections to Z + 2 jets production at the LHC }

\author{\speaker{Ansgar Denner}%
        \\
        Universit\"at W\"urzburg\\
        E-mail: \email{denner@physik.uni-wuerzburg.de}}

\author{Lars Hofer%
         \thanks{Present address: Universitat Aut\`onoma de Barcelona.}\\
                Universit\"at W\"urzburg\\
        E-mail: \email{lhofer@ifae.es}}
\author{Andreas Scharf\\
                Universit\"at W\"urzburg\\
        E-mail: \email{ascharf@physik.uni-wuerzburg.de}}
\author{Sandro Uccirati\\
        Universit\`a di Torino\\
        E-mail: \email{uccirati@to.infn.it}}

      \abstract{We present results for the electroweak radiative
        corrections to the production of a leptonically decaying Z
        boson in association with two jets at the LHC. Tree-level and
        one-loop amplitudes have been obtained with the computer code
        \recola\ for the recursive generation of tree-level and
        one-loop amplitudes in the Standard Model. The one-loop
        integrals have been calculated with the tensor-integral
        library \collier. The electroweak corrections turn out to be
        small for inclusive cross sections but can become large in
        tails of distributions.}

\FullConference{11th International Symposium on Radiative Corrections (Applications of Quantum Field Theory
to Phenomenology) \\
                 22-27 September 2013\\
                 Lumley Castle Hotel, Durham, UK}

\begin{document}

\section{Introduction}

The discovery of a Higgs boson with a mass of around $125\GeV$ was a
spectacular success of the \ac{SM}. Having completed the particle
spectrum of the \ac{SM}, the task is to scrutinise this model in all
details and to search for experimental signals showing deviations from
their \ac{SM} predictions. This includes the detailed
investigation of the observed Higgs boson as well as precise
measurements of all kind of \ac{SM} processes. For an adequate
comparison between theory and experiment higher-order perturbative
corrections are indispensable. This concerns in the first place QCD
corrections, which are typically of the order of several ten per cent
or more, and even \ac{NNLO} QCD corrections are needed for several
processes. However, also \ac{EW} corrections are mandatory for many
processes at the LHC. While they are often (but not always) small for
inclusive observables, they can be enhanced by kinematic effects, in
particular near resonances, or where high-energy scales matter, like
in tails of distributions. They can easily reach several ten per cent
for energy scales in the TeV region.

In the past years, huge progress has been made in the automation of
the calculation of \ac{NLO} QCD corrections. Different groups have
developed software packages \cite{Berger:2008sj, Giele:2008bc,
  Lazopoulos:2008ex, Giele:2009ui, Badger:2010nx, Hirschi:2011pa,
  Bevilacqua:2011xh, Cullen:2011ac} and used them to perform
complicated calculations. The efforts have concentrated mostly on QCD
corrections (examples can be found in these proceedings). Recently we
have constructed \recola, a generator of one-loop (and tree)
amplitudes in the full \ac{SM} \cite{Actis:2012qn,Uccirati_radcor},
including \ac{EW} corrections.
It is based on an algorithm which uses recursion relations for the
computation of the coefficients of the tensor integrals
\cite{vanHameren:2009vq}.

As a first application of \recola~we have chosen the process $\Pp\Pp \to
\PZ+2\,$jets$\,\to \Plp\Plm+2\,$jets.  Owing to its large cross section
and similar signatures it provides a major background for Higgs-boson
production in vector-boson fusion kinematics and allows to study the
systematics for the $\PH+2\,$jets final state. The dominant NLO QCD
corrections of $\O(\alphas^3\alpha^2)$ have been investigated in
\citeres{Campbell:2002tg}, while a subset of EW
$\PZ+2\,$jets production has been studied at \ac{NLO} QCD in
\citere{Oleari:2003tc}.  Electroweak Sudakov corrections have been
considered in \citere{Chiesa:2013yma}.
Our aim is the calculation of the complete \ac{EW}
corrections of $\O(\alphas^2\alpha^3)$.

\section{\collier, a Fortran library for tensor one-loop integrals}

A general one-loop amplitude $\de\M$ can be written as
\beqar
\delta\M &=&
\sum_{j}\,\sum_{R_j}\,c_{\mu_1\cdots\mu_{R_j}}^{(j,R_j,{N_j})}\,
T_{(j,R_j,{N_j})}^{\mu_1\cdots\mu_{R_j}},
\eeqar
where $j$ runs over all appearing tensor integrals with rank $R_j$ and
degree (number of propagators) $N_j$.

While {\recola} calculates the coefficients
$c_{\mu_1\cdots\mu_{R_j}}^{(j,R_j,{N_j})}$, 
the tensor integrals
\beq
T_{(j,R_j,{N_j})}^{\mu_1\cdots\mu_{R_j}} =
{\frac{(2\pi\mu)^{4-D}}{\ri\pi^2}}
\int
\rd^Dq\, \frac{q^{\mu_1} \cdots q^{\mu_{R_j}}}{D_{j,0}\cdots D_{j,{N_j-1}}},
\qquad
{D_{j,a} = (q+p_{j,a})^2 - m_{j,a}^2}
\eeq
are provided  by {\collier}, the Complex One-Loop LIbrary with Extended
Regularisations \cite{Denner:2002ii,Denner:2005nn,Denner:2010tr,collier}.
This library is also used by {\sc OpenLoops}
\cite{Cascioli:2011va,Maierhoefer:Racdor13}.

{\collier} is a Fortran library for the numerical evaluation of
one-loop scalar and tensor integrals appearing in perturbative
calculations in relativistic quantum field theory. It provides tensor
integrals with arbitrary rank for $N$-point functions up to currently $N=6$,
either directly as numerical values for the tensor elements
$T_{(j,R_j,{N_j})}^{\mu_1\cdots\mu_{R_j}}$ or as numerical values for
the invariant coefficients
$T^{(j,R_j,N_j)}_{\underbrace{\scriptstyle 0\ldots0}_{2n}
  i_{2n+1}\ldots i_{R_j}}$ appearing in the covariant decomposition
\beq
T_{(j,R_j,{N_j})}^{\mu_1\ldots\mu_{R_j}} =
\sum_{n=0}^{\left[\frac{R_j}{2}\right]} \,
\sum_{i_{2n+1},\ldots,i_{R_j}=1}^{N-1} \,
\{\underbrace{g \ldots g}_n  p\ldots p\}^{\mu_1\ldots\mu_{R_j}}_{i_{2n+1}\ldots i_{R_j}}
\, T^{(j,R_j,N_j)}_{\underbrace{\scriptstyle 0\ldots0}_{2n} i_{2n+1}\ldots i_{R_j}}.
\eeq

The tensor integrals are evaluated as follows: For 1-point and 2-point
tensor integrals explicit numerically stable expressions are used
\cite{Passarino:1978jh}. For 3-point and 4-point functions we employ
Passarino--Veltman reduction along with the tensor-integral reduction
methods of \citere{Denner:2005nn} which are based on expansions in
small determinants.  Depending on the specific kinematic configuration
a suitable method is selected for each tensor integral. In this way a
numerically stable calculation is achieved for almost all
configurations. Finally 5-point and 6-point tensor integrals are
directly reduced to integrals with lower rank and lower degree using
the methods summarised in \citeres{Denner:2002ii,Denner:2005nn}.

{\collier} includes a complete set of scalar integrals for scattering
processes.  These are evaluated from analytical expressions as given
in \citere{Denner:2010tr,'tHooft:1978xw}.  Ultraviolet singularities
are regularised dimensionally, soft and collinear singularities can be
regularised either dimensionally or with masses. While complex
internal masses, required for unstable particles, are fully supported,
external momenta and virtualities must be real.
 
{\collier} has a built-in cache system to avoid the recalculation of
identical integrals and two branches that allow for an independent
calculation of each integral via two different implementations and direct
numerical cross-checks.

\section{$\PZ+2\,$jets production at the LHC}
\label{ppZjj}

We consider the production of a leptonically decaying $\PZ$ boson in
association with 2 hard jets. We include diagrams including a resonant $\PZ$
boson as well as all irreducible background diagrams leading to the
$\Pl^+\Pl^-\Pj\Pj$ final state.

\subsection{Leading-order contributions}

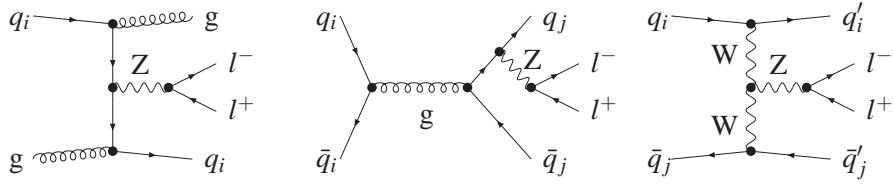
\begin{figure}
\centerline{
{\unitlength .6pt 
\begin{picture}(125,100)(0,0)
\SetScale{.6}
\ArrowLine( 15,95)( 65,90)
\Gluon( 15, 5)( 65,10){3}{7}
\ArrowLine( 65,10)(115, 5)
\Gluon(115,95)( 65,90){3}{7}
\ArrowLine( 65,90)( 65,40)
\ArrowLine( 65,40)( 65,10)
\Photon(100,50)( 65,50){3}{4}
\ArrowLine( 100,50)( 130,65)
\ArrowLine( 130,35)( 100,50)
\Vertex(65,90){3}
\Vertex(65,10){3}
\Vertex(65,50){3}
\Vertex(100,50){3}
\put(  0,90){$\Pq_i$}
\put(  0,0){$\Pg$}
\put(123,90){{$\Pg$}}
\put(123,0){{$\Pq_i$}}
\put(77,59){{$\PZ$}}
\put(138,60){{$\Plm$}}
\put(138,30){{$\Plp$}}
\SetScale{1}
\end{picture}
}
\hspace*{3em}
{\unitlength .6pt 
\begin{picture}(145,100)(0,0)
\SetScale{.6}
\ArrowLine( 15,95)( 35,50)
\ArrowLine( 35,50)( 15, 5)
\ArrowLine(95,50)(115,72.5)
\ArrowLine(115,72.5)(135,95)
\ArrowLine(135, 5)(95,50)
\Gluon( 35,50)(95,50){3}{7}
\Photon(115,72.5)(135,50){3}{4}
\ArrowLine( 135,50)( 165,65)
\ArrowLine( 165,35)( 135,50)
\Vertex( 35,50){3}
\Vertex(95,50){3}
\Vertex( 115,72.5){3}
\Vertex(135,50){3}
\put(  0,90){$q_i$}
\put(  0,0){$\bar{q}_i$}
\put(143,90){{$\Pq_j$}}
\put(143,0){{$\bar{\Pq}_j$}}
\put(173,60){{$\Plm$}}
\put(173,30){{$\Plp$}}
\put(131,62){{$\PZ$}}
\put( 65,28){{$\Pg$}}
\SetScale{1}
\end{picture}
\hspace*{3em}
{\unitlength .6pt 
\begin{picture}(125,100)(0,0)
\SetScale{.6}
\ArrowLine( 15,95)( 65,90)
\ArrowLine( 65,10)( 15, 5)
\ArrowLine(115, 5)( 65,10)
\ArrowLine( 65,90)(115,95)
\Photon( 65,10)( 65,90){3}{7}
\Photon(100,50)( 65,50){3}{4}
\ArrowLine( 100,50)( 130,65)
\ArrowLine( 130,35)( 100,50)
\Vertex(65,90){3}
\Vertex(65,10){3}
\Vertex(65,50){3}
\Vertex(100,50){3}
\put(  0,90){$q_i$}
\put(  0,0){$\bar{\Pq}_j$}
\put(123,90){{$q'_i$}}
\put(123,0){{$\bar{\Pq}'_j$}}
\put(138,60){{$\Plm$}}
\put(138,30){{$\Plp$}}
\put( 77,59){{$\PZ$}}
\put( 40,65){{$\PW$}}
\put( 40,20){{$\PW$}}
\SetScale{1}
\end{picture}
}
}
}
\caption{From left to right: Sample tree diagrams for the QCD
  contributions to $\Pq_i\,\Pg\to\Pq_i\,\Pg\,\Plp\,\Plm$ and to
  $\Pq_i\,\bar{\Pq}_i\to\Pq_j\,\bar{\Pq}_j\,\Plp\,\Plm$, and the EW
  contributions to $\Pq_i\,\bar{\Pq}_j\to\Pq'_i\,\bar{\Pq}_j'\,\Plp\,\Plm$.} 
\label{fig:LOdiags}
\end{figure}

At \ac{LO}, the production of a leptonically decaying $\PZ$~boson at
the LHC in association with a pair of hard jets is governed by the
partonic subprocesses
\begin{eqnarray}
   \Pq_i\,\Pg&\to&\Pq_i\,\Pg\,\Plp\,\Plm\,,\label{eq:born_qg}\\
   \Pq_i\,\bar{\Pq}_i&\to&\Pq_j\,\bar{\Pq}_j\Plp\,\Plm\,,
   {\qquad q_i,q_j=\Pu,\Pc,\Pd,\Ps,\Pb}
\label{eq:born_qq},
\end{eqnarray}
and their crossing-related counterparts.  All partonic processes can
be constructed from the three basic channels $\Pu\Pg\to
\Pu\Pg\Plp\Plm$, $\Pu\Ps\to \Pu\Ps\Plp\Plm$, and $\Pu\Ps\to
\Pd\Pc\Plp\Plm$.
While the mixed quark--gluon (gluonic) channels \refeq{eq:born_qg}
contribute to the cross section exclusively at order
$\ord(\alphas^2\alpha^2)$, the four-quark channels \refeq{eq:born_qq}
involve LO diagrams of strong as well as of EW nature leading to
contributions of order $\ord(\alphas^2\alpha^2)$,
$\ord(\alphas\alpha^3)$, and $\ord(\alpha^4)$. Representative Feynman
diagrams are shown in \reffi{fig:LOdiags}.  Photon induced
contributions, which are below $0.05\%$, have been omitted.


\subsection{Setup of \ac{NLO} calculation}

Taking into account all one-loop QCD and EW diagrams, the cross
section involves contributions of orders $\ord(\alphas^3\al^2)$,
$\ord(\alphas^2\al^3)$, $\ord(\alphas\al^4)$, and $\ord(\al^5)$. We
are interested in the contributions of $\ord(\alphas^2\al^3)$ which
involve \ac{NLO} \ac{EW} corrections to the dominant \ac{LO}
contributions of $\ord(\alphas^2\al^2)$ and \ac{NLO} QCD corrections
to the LO interference contributions of $\ord(\alphas\al^3)$.

We use the complex-mass scheme
\cite{Denner:1999gp,Denner:2005fg,Denner:2006ic} to treat resonant
$\PZ$-boson propagators
(see \reffi{fig:NLOdiags} left for a sample diagram), 
\ie we consistently use the complex masses
\begin{equation}
\mu_{\PZ}^2 = \MZ^2 -\ri \MZ \GZ, \qquad
\mu_{\PW}^2 = \MW^2 -\ri \MW \GW
\end{equation}
for the $\PZ$ and $\PW$ bosons.  
   
The electromagnetic coupling constant $\alpha$ is defined within
the $G_\mu$ scheme, \ie we use
\beq\label{eq:Gmu}
 \alpha = \frac{\sqrt{2}\GF\MW^2}{\pi}\left(1-\frac{\MW^2}{\MZ^2}\right).
\eeq
This definition takes into account higher-order effects of the
renormalisation-group running from $0$ to $\MW^2$. Moreover,
the renormalisation of $\alpha$ becomes independent of light quark masses.

\subsection{Virtual corrections}
\begin{figure}
\centerline{
{\unitlength .8pt 
\begin{picture}(125,110)(0,0)
\SetScale{.8}
\Gluon( 15,95)( 55,75){3}{5}
\Gluon( 15, 5)( 55,25){3}{5}
\ArrowLine(55,75)(55,25)
\ArrowLine(55,25)(95,25)
\ArrowLine(95,25)(95,75)
\ArrowLine(95,75)(55,75)
\Photon(95,25)(135,25){3}{4}
\ArrowLine(135,95)(165,75)
\ArrowLine(165,115)(135,95)
\Photon(95,75)( 135,95){3}{4}
\ArrowLine(135,25)(165,5)
\ArrowLine(165,45)(135,25)
\Vertex(55,25){3}
\Vertex(55,75){3}
\Vertex(95,25){3}
\Vertex(95,75){3}
\Vertex(135,25){3}
\Vertex(135,95){3}
\put(-3,90){$\rm g$}
\put(-3,0){$\rm g$}
\put(113,100){{$\rm Z$}}
\put(173,0){{$q_i$}}
\put(173,40){{$\bar{q}_i$}}
\put(173,70){{$l^-$}}
\put(173,110){{$l^+$}}
\put(113,5){{{$\rm Z$}}}
\SetScale{1}
\end{picture}
}
\hspace*{9em}
{\unitlength .8pt 
\begin{picture}(160,120)(0,0)
\SetScale{.8}
\ArrowLine( 15,95)( 45,90)
\Gluon( 45,10)( 15, 5){3}{3}
\ArrowLine( 95,90)( 125,95)
\Gluon( 125,5)(95, 10){3}{3}
\ArrowLine(45,90)(45,10)
\ArrowLine(45,10)(95,10)
\ArrowLine(95,10)(130,30)
\ArrowLine(130,70)(95,90)
\ArrowLine(165,70)(130,70)
\ArrowLine(135,30)(165,30)
\Photon(45,90)(95,90){3}{7}
\Photon(130,70)( 130,30){3}{4}
\Vertex(45,90){3}
\Vertex(45,10){3}
\Vertex(95,90){3}
\Vertex(95,10){3}
\Vertex(130,70){3}
\Vertex(130,30){3}
\put(-3,90){$q_i$}
\put(-3,0){$\rm g$}
\put(168,30){{$q_i$}}
\put(128,0){{$\rm g$}}
\put(137,45){{$\ga,\PZ,\PW$}}
\put(50,100){{$\ga,\PZ,\PW$}}
\put(168,70){{$l^+$}}
\put(128,90){{$l^-$}}
\SetScale{1}
\end{picture}
}
}
\caption{Box diagram involving potentially resonant
  $\PZ$-boson propagators (left) and hexagon diagram involving a
  6-point tensor integral of rank 4 (right).}
\label{fig:NLOdiags}
\end{figure}
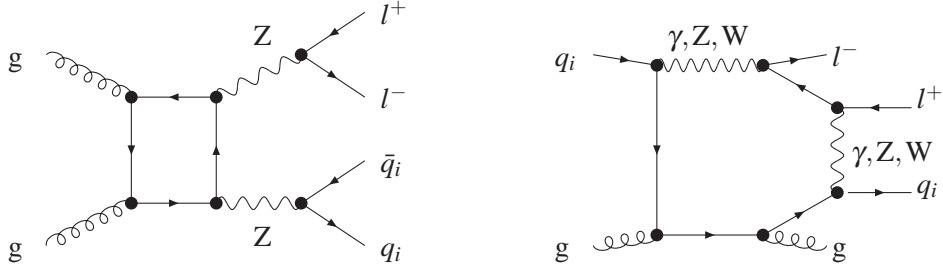

The virtual corrections contributing at $\ord(\alphas^2\alpha^3)$
involve $\ord(1200)$ diagrams for the $\Pu\Pg\to\Pu\Pg\Plp\Plm$
channel, including 18 hexagons and 85 pentagons, and an almost
comparable number of diagrams for the $\Pu\Ps\to\Pu\Ps\Plp\Plm$
channel.
The most complicated topologies involve 6-point functions up to rank
$4$ (see \reffi{fig:NLOdiags} right for a sample diagram).
Renormalisation is performed within the complex-mass scheme
\cite{Denner:2005fg}.  As the coupling $\alpha_{G_\mu}$ is derived
from $\MW$, $\MZ$ and $G_\mu$, its counterterm inherits a correction
term $\Delta r$ from the weak corrections to muon decay.

In NLO we neglect all channels involving external bottom quarks.
At LO these contributions are taken into account and contribute at the
per-cent level.

\subsection{Real corrections}
\label{real_corr}

The real EW corrections of $\ord(\alphas^2\alpha^3)$ consist of real
photon emission from the LO QCD diagrams  in all subprocesses
\refeq{eq:born_qg}, \ie the processes
\beqar
   \Pq_i\,\Pg&\to&\Pq_i\,\Pg\,\Plp\,\Plm\,\gamma\,,\\ \label{eq:real_qg}
   \Pq_i\,\bar{\Pq}_i&\to&\Pq_j\,\bar{\Pq}_j\Plp\,\Plm\,\ga\,,
\eeqar
and from real gluon emission contributions in the interferences
between LO QCD and EW diagrams, \ie the processes
\beq
   \Pq_i\,\bar{\Pq}_i \to\Pq_j\,\bar{\Pq}_j\Plp\,\Plm\,\Pg\,.
\eeq
The amplitudes can be constructed in exactly the same way as for the
LO, but we only take contributions of  $\ord(\alphas^2\alpha^3)$ into
account.

IR divergences of soft or collinear photons and gluons are regularised
dimensionally. We use the Catani--Seymour dipole formalism as formulated in
\citere{Catani:1996vz}, which we transferred in a straightforward way
to the case of dimensionally regularised photon emission.

The presence of gluons in the final state gives rise to a further type
of singularity. In IR-safe observables quarks, and thus all QCD
partons, have to be recombined with photons if they are sufficiently
collinear. Thus, soft gluons can pass the selection cuts if they are
recombined with a hard collinear photon, giving rise to a soft-gluon
divergence that would be cancelled by the virtual QCD corrections to
$\PZ+1\,\mathrm{jet}+\gamma$ production.  Following
\citeres{Denner:2010ia} we eliminate this singularity by
discarding events containing a jet consisting of a hard photon and a
soft parton $a$ ($a=q_i,\bar{q}_i,\Pg$) taking the photon--jet energy
fraction $z_{\gamma}= E_{\gamma}/(E_{\gamma}+E_{a})$ as a
discriminator.  Finally, we have to absorb left-over singularities into
contributions involving the quark--photon fragmentation function
\cite{Glover:1993xc}.


\subsection{Implementation}

The calculation was performed using the amplitude generator \recola\ 
for all tree-level, one-loop and bremsstrahlung amplitudes.  This
generator is interfaced with the tensor-integral library {\sc Collier}
\cite{collier}. The phase-space integration is performed by
an in-house multi-channel Monte-Carlo generator \cite{Motz}.

The results have been checked by a second independent calculation
based on the conventional Feynman-diagrammatic approach. It uses {\sc
  FeynArts}~3.2 \cite{Hahn:2000kx,Hahn:2001rv}, {\sc FormCalc}~3.1
\cite{Hahn:1998yk} and {\sc Pole} \cite{Accomando:2005ra} for the
generation, simplification and calculation of the Feynman amplitudes.
The tensor-integral coefficients are again evaluated by {\sc Collier},
which includes a second independent implementation of all its building
blocks. Finally, the phase-space integration is performed with the
multi-channel generator {\sc Lusifer} \cite{Dittmaier:2002ap}.

\subsection{Numerical results}
\label{ppZjj results}

We present results for the 8TeV LHC with the input parameters as given
in \citere{Actis:2012qn}.  Since we focus on NLO EW corrections, we
stick to the LO MSTW2008LO PDF set \cite{Martin:2009iq}.  For the QCD
factorisation and renormalisation scales we choose
$\mu_{\mathrm{F}}=\mu_{\mathrm{R}}=\MZ$.  The scale choice as well as
the actual value for the strong coupling $\alphas$ plays a minor role
for our numerical analysis of EW radiative corrections.

The jet clustering is performed with the anti-$k_\mathrm{T}$
algorithm \cite{Cacciari:2008gp} with separation parameter $R=0.4$.
We require two hard jets and two charged leptons fulfilling the
following requirements
\beqar
   \label{eq:cuts}
   p_{\mathrm{T},\mathrm{jet}}>30\GeV, \qquad |y_\mathrm{jet}|<4.5, \qquad
   p_{\mathrm{T},\Pl}>20\GeV, \qquad |y_\Pl|<2.5,\nl
\De R_{\Pl\Pl}>0.2, \qquad \De R_{\Pl\mathrm{jet}}>0.5, \qquad
66\GeV<M_{\Pl\Pl}<116\GeV.
\eeqar
for the transverse momenta, rapidities, rapidity--azimuthal angle
separation, and invariant mass.  In addition, the photon energy
fraction $z_\ga$ in a jet must be below $0.7$.

Using the set-up defined above we find for the cross
section the results shown in \refta{tab:x-section}.
\begin{table}
\centerline{
   \begin{tabular}{|c|c|c|c|c|}
     \hline
&&&&\\[-.4cm]
process class  & $\sigma^\LO~ [\rm{ fb}] $  & $ \sigma^\LO/\sigma^\LO_{\rm tot}$ $[\%]$   & $\sigma_\EW^\NLO~ [\rm{ fb}] $  & $ \frac{\sigma_\EW^\NLO}{\sigma^\LO}-1$ $[\%]$ \\[+.2cm]
\hline
     \hline gluonic &  17948(4)    &  77.3 & 17534(4)   & {-2.31}  \\
     \hline four-quark & 5270.0(5) & 22.7  & 5140(2)  & {-2.46}  \\
     \hline sum & 23218(4)  & 100 & 22675(5) & {-2.34} \\
     \hline
   \end{tabular}}
  \caption{LO cross section for 
    $\Pp\Pp \to \Plp\Plm+2\,$jets at the $8\TeV$ LHC split into
    contributions from gluonic and four-quark channels. The second
    column provides the LO cross section with integration error on the
    last digit in parentheses, the third column contains the relative
    contribution to the total cross section in per cent, the fourth
    column the NLO EW cross section, and in the last column the
    relative EW corrections.}
  \label{tab:x-section}
\end{table}
The total cross section is dominated by processes with external
gluons, which amount to 77\%. For our set of cuts, the corresponding
relative EW corrections are at the level of $-2.5\%$ and similar for
all channels.

\begin{figure}
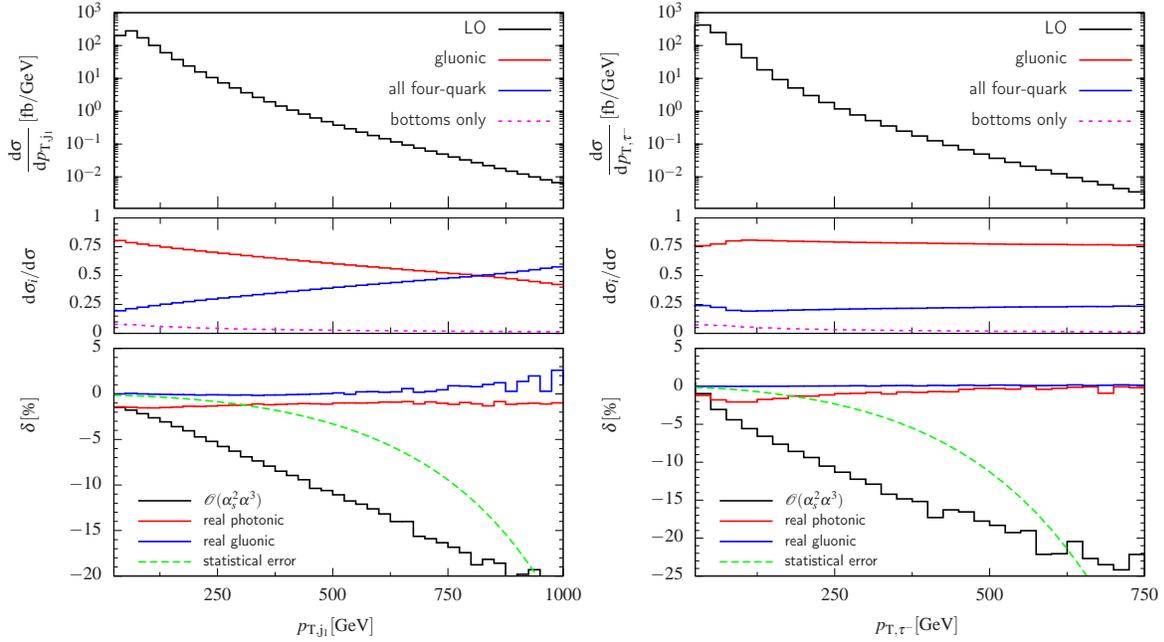

\centerline{\hspace*{3em}
{\scalebox{0.48}{\input{LO_LHC8_pTjetH.tex}}}\hspace{10em}
{\scalebox{0.48}{\input{LO_LHC8_pTTauM.tex}}}}
\strut\\[-3ex]
\centerline{\hspace*{3em}
{\scalebox{0.48}{\input{NLO_LHC8_pTjetH_rel.tex}}}\hspace{4em}
{\scalebox{0.48}{\input{NLO_LHC8_pTTauM_rel.tex}}}}
\strut\\[-9ex]
\caption{Distributions of the transverse momenta  
  of the harder jet ${\Pj_1}$ (left) and the negatively charged lepton
  ${\Plm}$ (right) at the $8\TeV$ LHC at LO (black). The central panels
  show the relative contributions of gluonic (red), four-quark (blue)
  and bottom (magenta, dashed) channels. In the lower panels we show
  the total relative corrections (black) from $\ord(\alphas^2\alpha^3)$
  contributions, the included real photonic (red) and real gluonic (blue)
  corrections together with the statistical error (green, dashed).}
\label{fig:num1}
\end{figure}
In \reffi{fig:num1} we present results for the differential cross
section as a function of the transverse momentum of the harder jet and of
the negatively charged lepton. Both distributions decrease over five
orders of magnitude in the displayed $\pT$ ranges. While the
four-quark channels dominate for high $p_{\rT,\Pj_1}$, their
contribution stays at the level of $25\%$ for all $p_{\rT,\Plm}$. The
bottom-quark contributions, which are shown separately, are below
$5\%$ and become less important for large transverse momenta. The
electroweak corrections are negative and reach about $-20\%$ for
$p_{\rT,\Pj_1}\approx1\TeV$ and $p_{\rT,\Plm}\approx600\GeV$. The
large negative corrections result from Sudakov logarithms in the EW
loop contributions. The (subtracted) real corrections from photon or
gluon emission stay below $2\%$. We also show the experimental
statistical uncertainty corresponding to an integrated luminosity of
$20\fba^{-1}$.

\section{Conclusions}
\label{conclusions}

We have presented results for the $\ord(\alphas^2\alpha^3)$ contributions to
the production of $\Plp\Plm+2\,$jets at the LHC. These include
electroweak corrections to the LO QCD diagrams and QCD
corrections to the interferences between LO QCD and EW
diagrams. The results have been obtained with the amplitude generator
\recola\ and the tensor-integral library \collier. The
electroweak corrections turn out to be at the level of $-2\%$ for
inclusive cross sections. In the high-energy tails of distributions
large EW corrections appear which can be
attributed to EW Sudakov logarithms.

\subsection*{Acknowledgements}
This work was supported in part by 
the Deutsche
Forschungsgemeinschaft (DFG) under reference number DE~623/2-1.


\begin{thebibliography}{99}

\bibitem{Berger:2008sj}
  C.~F.~Berger, \ea,
  Phys.\ Rev.\ D {\bf 78} (2008) 036003
  [arXiv:0803.4180 [hep-ph]].

  \bibitem{Giele:2008bc}
  W.~T.~Giele and G.~Zanderighi,
  JHEP {\bf 0806} (2008) 038
  [arXiv:0805.2152 [hep-ph]].

  \bibitem{Lazopoulos:2008ex}
  A.~Lazopoulos,
  arXiv:0812.2998 [hep-ph].

  \bibitem{Giele:2009ui}
  W.~Giele, Z.~Kunszt and J.~Winter,
  Nucl.\ Phys.\ B {\bf 840} (2010) 214
  [arXiv:0911.1962 [hep-ph]].

\bibitem{Badger:2010nx}
  S.~Badger, B.~Biedermann and P.~Uwer,
  Comput.\ Phys.\ Commun.\  {\bf 182} (2011) 1674
  [arXiv:1011.2900 [hep-ph]].

\bibitem{Hirschi:2011pa}
  V.~Hirschi, \ea,
  JHEP {\bf 1105} (2011) 044
  [arXiv:1103.0621 [hep-ph]].

\bibitem{Bevilacqua:2011xh}
  G.~Bevilacqua, \ea,
  arXiv:1110.1499 [hep-ph].

  \bibitem{Cullen:2011ac}
  G.~Cullen, \ea,
  Eur.\ Phys.\ J.\ C {\bf 72} (2012) 1889
  [arXiv:1111.2034 [hep-ph]].

\bibitem{Actis:2012qn}
  S.~Actis, \ea,
  JHEP {\bf 1304} (2013) 037
  [arXiv:1211.6316 [hep-ph]].

\bibitem{Uccirati_radcor}
  S.~Actis, A.~Denner, L.~Hofer, A.~Scharf and S.~Uccirati, these proceedings.

\bibitem{vanHameren:2009vq}
  A.~van Hameren,
  JHEP {\bf 0907} (2009) 088
  [arXiv:0905.1005 [hep-ph]].

\bibitem{Campbell:2002tg}
  J.~M.~Campbell and R.~K.~Ellis,
  Phys.\ Rev.\ D {\bf 65} (2002) 113007
  [hep-ph/0202176];\\
%
  J.~M.~Campbell, R.~K.~Ellis and D.~L.~Rainwater,
  Phys.\ Rev.\ D {\bf 68} (2003) 094021
  [hep-ph/0308195].

\bibitem{Oleari:2003tc}
  C.~Oleari and D.~Zeppenfeld,
  Phys.\ Rev.\ D {\bf 69} (2004) 093004
  [hep-ph/0310156].

\bibitem{Chiesa:2013yma}
M.~Chiesa, \ea,
  Phys.\ Rev.\ Lett.\  {\bf 111} (2013) 121801
  [arXiv:1305.6837 [hep-ph]].

\bibitem{Denner:2002ii}
  A.~Denner and S.~Dittmaier,
  Nucl.\ Phys.\ B {\bf 658} (2003) 175
  [hep-ph/0212259].

\bibitem{Denner:2005nn}
  A.~Denner and S.~Dittmaier,
  Nucl.\ Phys.\ B {\bf 734} (2006) 62
  [hep-ph/0509141].

\bibitem{Denner:2010tr}
  A.~Denner and S.~Dittmaier,
  Nucl.\ Phys.\ B {\bf 844} (2011) 199
  [arXiv:1005.2076 [hep-ph]].

\bibitem{collier}
  A.~Denner, S.~Dittmaier and L.~Hofer,
  in preparation.

\bibitem{Cascioli:2011va}
  F.~Cascioli, P.~Maierh\"ofer and S.~Pozzorini,
  Phys.\ Rev.\ Lett.\  {\bf 108} (2012) 111601
  [arXiv:1111.5206 [hep-ph]].

\bibitem{Maierhoefer:Racdor13}
  F.~Cascioli, P.~Maierh\"ofer and S.~Pozzorini, these proceedings.

\bibitem{Passarino:1978jh}
  G.~Passarino and M.~J.~G.~Veltman,
  Nucl.\ Phys.\ B {\bf 160} (1979) 151.

\bibitem{'tHooft:1978xw}
  G.~'t Hooft and M.~J.~G.~Veltman,
  Nucl.\ Phys.\ B {\bf 153} (1979) 365.

\bibitem{Denner:1999gp}
  A.~Denner, S.~Dittmaier, M.~Roth and D.~Wackeroth,
  Nucl.\ Phys.\ B {\bf 560} (1999) 33
  [hep-ph/9904472].

\bibitem{Denner:2005fg}
  A.~Denner, S.~Dittmaier, M.~Roth and L.~H.~Wieders,
  Nucl.\ Phys.\ B {\bf 724} (2005) 247
   [Erratum-ibid.\ B {\bf 854} (2012) 504]
  [hep-ph/0505042].

\bibitem{Denner:2006ic}
  A.~Denner and S.~Dittmaier,
  Nucl.\ Phys.\ Proc.\ Suppl.\  {\bf 160} (2006) 22
  [hep-ph/0605312].

\bibitem{Catani:1996vz}
  S.~Catani and M.~H.~Seymour,
  Nucl.\ Phys.\  B {\bf 485} (1997) 291
  [Erratum-ibid.\  B {\bf 510} (1998) 503]
  [hep-ph/9605323].

\bibitem{Denner:2010ia}
  A.~Denner, S.~Dittmaier, T.~Gehrmann and C.~Kurz,
  Nucl.\ Phys.\  {\bf B836} (2010)  37 
  [arXiv:1003.0986 [hep-ph]].

\bibitem{Glover:1993xc}
  E.~W.~N.~Glover and A.~G.~Morgan,
  Z.\ Phys.\  C {\bf 62} (1994) 311;\\
%
  D.~Buskulic {\it et al.}  [ALEPH Collaboration],
  Z.\ Phys.\  C {\bf 69} (1996) 365.

\bibitem{Motz}
  T.~Motz, PhD thesis, Z\"urich 2011.

\bibitem{Hahn:2000kx}
  T.~Hahn,
  Comput.\ Phys.\ Commun.\  {\bf 140} (2001) 418
  [hep-ph/0012260].

\bibitem{Hahn:2001rv}
  T.~Hahn and C.~Schappacher,
  Comput.\ Phys.\ Commun.\  {\bf 143} (2002) 54
  [hep-ph/0105349].

\bibitem{Hahn:1998yk}
  T.~Hahn and M.~P\'erez-Victoria,
  Comput.\ Phys.\ Commun.\  {\bf 118} (1999) 153
  [hep-ph/9807565].

\bibitem{Accomando:2005ra}
  E.~Accomando, A.~Denner and C.~Meier,
  Eur.\ Phys.\ J.\  C {\bf 47} (2006) 125
  [hep-ph/0509234].

\bibitem{Dittmaier:2002ap}
  S.~Dittmaier and M.~Roth,
  Nucl.\ Phys.\  B {\bf 642} (2002) 307
  [hep-ph/0206070].

\bibitem{Martin:2009iq}
  A.~D.~Martin {\it et al.},
  Eur.\ Phys.\ J.\  C {\bf 63}, (2009) 189
  [arXiv:0901.0002 [hep-ph]].

\bibitem{Cacciari:2008gp}
  M.~Cacciari, G.~P.~Salam and G.~Soyez,
  JHEP {\bf 0804} (2008) 063
  [arXiv:0802.1189 [hep-ph]].



\end{thebibliography}
\end{document}